**A Hibonite-Pyroxene Spherule in Allan Hills 77307 (CO3.03): Petrography and Mineralogy**

Ritesh Kumar Mishra[1, 2*]

[1]Center for Isotope Cosmochemistry and Geochronology,

Astromaterials Research and Exploration Science division,

EISD-XI, NASA-Johnson Space Center,

 2101, NASA Parkway, Houston, TX 77058, USA,

now at [2*]Institut fur Geoweissenchaften, Im Neuenheimer Feld 234-236

Ruprecht-Karls-Universität, Heidelberg, D 69210 Germany.

email: riteshkumarmishra@gmail.com

Ph:+49-6221-544825

**Abstract**

Hibonite-pyroxene spherules are an extremely rare kind of refractory inclusion that show a wide range of "exotic" isotopic properties despite their defining similarity and simplicity in morphology and mineralogy. One such, relatively large (~120μm diameter), inclusion has been found in one of the most pristine meteorites, Allan Hills 77307 (a carbonaceous chondrite of the Ornans group; Petrologic type 3.03). The inclusion consists of two central hibonite laths of ~ 30×15 μm surrounded by Al, Ca-rich pyroxene. The hibonite laths have uniform composition. The composition of pyroxene surrounding the hibonite is radially homogenously Al,-Ca rich up to ~50-60 microns which transitions to Mg, -Ti rich at the outer boundary. Hibonite-pyroxene spherule found in ALHA 77307 shares many similarities with the other previously found hibonite-pyroxene spherules. A distinguishing feature of the inclusion in ALHA77307 is the presence of two slivers/ wedges at the opposite outer edge of the hibonite- pyroxene spherule that consist of rapidly, poorly crystalized, sub-micron minerals with pristine textures. The pristine petrography and mineralogy of this inclusion allow discernment of the expected general trend of formation and alteration amongst hibonite-pyroxene spherules.

**Introduction: Early Solar System Evolution Records in Meteorites and Hibonite-Pyroxene Spherules**

Records of the earliest events and processes during the birth and early evolution of the Solar system are preserved in various components of meteorites. The first forming solids of the Solar system- so called, Calcium, -Aluminium-rich inclusions (CAI), date the formation of the Solar system at ~ (4567.2 ± 0.5) Ma (Amelin et al. 2010; Bouvier and Wadhwa 2010). CAI is a generic name for a diverse suite of inclusions in chondrites that are constituted by refractory oxides and silicates mostly of calcium, aluminium, and a few other less refractory elements like titanium, magnesium, and iron. In an oxidizing environment (C/O <1) of a gas of solar composition at a total pressure of <$10^{-3}$ bar, model calculations predict formation of corundum ($Al_2O_3$), hibonite ($CaAl_{12}O_{19}$), grossite ($CaAl_4O_7$), perovskite ($CaTiO_3$), melilite (Gehlenite $CaAl_2SiO_7$-$CaMg_2Si_2O_7$ Akermanite), spinel ($MgAl_2O_4$), Ca-pyroxene, feldspar, olivine (($Fe,Mg)_2SiO_4$) sequentially under equilibrium conditions (Yoneda and Grossman 1995; Ebel 2006). These minerals in distinctive combinations governed by diverse petrogenetic settings have been observed in CAIs that are found in various chondrite types (MacPherson 2014; Krot et al. 2014; Scott and Krot 2014). The abundances and coexistence of these phases provide useful information to constrain critical parameters of the prevailing cosmochemical environment such as pathways, duration, timescales, cooling rates, fugacity etc.. Petrographic, mineralogical, and isotopic evidences in various components of meteorites display the expected local, transient, quasi state of equilibrium during formation or alternatively, demonstrate that later events obliterated the original pristine records. Hence, there is preponderance of some of these minerals (perovskite, spinel, melilite) while other like corundum, grossite ($CaAl_4O_7$), krotite ($CaAl_2O_4$) are rare (MacPherson 2014; Scott and Krot 2014). It is important to emphasis that modal abundances, mean sizes, and the typical mineralogies of refractory inclusions vary quite significantly even amongst different groups of carbonaceous chondrite (e.g. CV, CR, CO) and are distinct for each group (MacPherson 2014; Scott and Krot 2014). Notwithstanding the compositional differences, the different petrogenetic settings of mineralogical associations/assemblages and their petrographic fabric provide crucial information about their formation and thereby the prevalent cosmochemical conditions in the early Solar system.

**Rare abundance of hibonite bearing CAIs and hibonite- silicate inclusions**

Amongst the three classes of chondrites, Carbonaceous chondrites class (except CI type) have the highest abundance of CAIs (<0.01 CI - 3 volume % CV). CAIs within nine groups of carbonaceous chondrites (e.g., Vigarano (CV), Renazzo (CR), Ornans (CO), Mighei (CM), Ivuna (CI)) also display the widest range in mean size, mineralogy, morphological types and isotopic characteristics (MacPherson 2014; Scott and Krot 2014). Melilite, spinel, anorthite, and pyroxene are the most common and abundant constituent of these CAIs. The earliest condensates of oxides of calcium and aluminium (corundum, hibonite, krotite, grossite) are rare except for ubiquitous spinel and perovskite compared to silicates of these elements (Al, Ca), along with Mg, and Fe (MacPherson, 2014). Petrographic and isotopic records provided by hibonite bearing inclusions are important because hibonite is the $2^{nd}$ phase to condense at equilibrium following corundum. Therefore, hibonites record an earlier and higher temperature events compared to the quintessential melilite bearing (Type B/A) CAIs. A rare class of distinct objects called hibonite-pyroxene spherules were identified and systematically studied by Ireland et al. (1991). **Hi**bonite-Pyro**x**ene **S**pherule (hereafter referred as **HIXs**) are called so because of distinctive presence of hibonite laths, positioned mostly centrally, and surrounded by Ca, Al- rich pyroxene in a spherical shape. Since the descriptive work by Ireland et al. (1991) only five additional such objects, for a total inventory of 9 HIXs, have been reported in the literature (Russell et al. 1998; Simon et al. 1998; Guan et al. 2000). Their importance for early solar system studies emanates from their unique exotic characteristics: (1) large anomalies in $\delta^{48}$Ca (-45 to +40‰ (permil)) and $\delta^{50}$Ti (-60 to +105‰) (2) distinct rare earth elements (REE) abundance patterns in hibonite and surrounding pyroxene, with most hibonites showing 'type II' akin pattern indicative of formation of hibonite by fractional condensation (3) all HIXs (except for Y 791717) are devoid of clearly resolvable radiogenic excess in $\delta^{26}$Mg in hibonite suggestive of their formation in an environment which were either devoid of or with very low abundance of $^{26}$Al/$^{27}$Al of $\leq 1 \times 10^{-6}$ (4) most show well resolved deficits in $\delta^{25}$Mg (Kurat 1975; Grossman et al. 1988; Ireland et al. 1988, 1990, 1991; Tomeoka et al. 1992; Russell et al. 1998; Simon et al. 1998; Guan et al. 2000). The petrographic and isotopic evidences taken together suggest that HIXs record some of the earliest high temperature events of the evolving cosmochemical environment at the time of homogenization of the solar system.

Hibonite-pyroxene spherules show a genetic relation ship with other hibonite bearing/ rich CAIs from Murchison and Murray (carbonaceous Mighei type) meteorite. Hibonite bearing CAIs in Murchison (CM type) based on the mineralogy and morphology of the refractory inclusions were classified into Platy hibonite crystals (PLACs), Spinel hibonite inclusions (SHIBs), and Blue Aggregates (BAGs) (Ireland 1988). PLACs show the largest range in anomalies of $\delta^{48}$Ca (-60 to 100‰), $\delta^{50}$Ti (-70 to ~280‰) and a range of oxygen isotopic composition $\Delta^{17}$O from -28‰ to -17‰ with lower abundances of $^{26}$Al/$^{27}$Al from unresolved excesses to $(1-1.5)\times 10^{-5}$ (Ireland 1990; Sahijpal et al. 2000; Kööp et al. 2016a). On the hand SHIBs show uniform oxygen isotopic composition $\Delta^{17}O$ of -23‰ with near canonical $^{26}Al/^{27}Al$ abundances of ~5-1×10$^{-5}$ and much smaller anomalies in $\delta^{50}Ti$ (< 5‰) (Ireland 1990; Liu et al. 2009; Kööp et al. 2016b). These recent isotopic studies of oxygen, calcium, titanium and radiogenic nuclides ($^{26}Al$-$^{26}Mg$) in these objects provide records of homogenization, evolution of oxygen isotopes, and early solar system events. Hibonite-pyroxene spherules, as enumerated earlier, show isotopic properties that are in between PLACs and SHIBs. Hence, despite their phylogentic and mineralogical similarities and relatively small size, these objects form a very rare suite of objects that presented the extreme range of exotic isotopic, mineralogical properties that could help understand the early solar system event processes by forming a cog in the continuum from platy hibonite crystals (PLACs), spinel- hibonites (SHIBs), fractionation and unexplained nuclear effects (F/UN) CAIs to normal CAIs. These rare and exotic objects therefore provide records of unique situations/conditions in the early Solar system prior to the end of complete homogenization recorded by Type B CAIs (Mishra and Chaussidon 2014a; Chaussidon and Liu 2015). Table 1 lists some features of these nine hibonite-pyroxene spherules found so far. Four HIXs have been found in carbonaceous chondrites of the 'Ornans' type; one each in Colony (CO3.1), Yamato (Y) 791717 (CO3.3), Allan Hills (ALH) 82101 (CO3.4), Lancé (CO3.5) (Fig. S1). Three others have been found in carbonaceous chondrites of the Mighei type; two in Murchison (CM2) (MUR7-228, MUR 7-753), and one in Murray (CM2) (MYSM3) and one has been found in Allan Hills (ALH) 85085 (CH3) and another one in the enstatite chondrite Elephant Moraine (EET) 87746 (EH4) (Kurat 1975; Grossman et al. 1988; Ireland et al. 1991; Tomeoka et al. 1992; Russell et al. 1998; Simon et al. 1998; Guan et al. 2000). In this paper we report on the petrography and mineralogy of a hibonite-pyroxene spherule found in Allan Hills (ALH) A77307 (CO3.03). Notwithstanding the rarity of these objects, the identified and studied hibonite-pyroxene spherule in ALHA77307 presented in this paper has been found in one of the least altered meteorite (3.03) available in our collection, providing a distinct possibility of best preserved petrographic and isotopic records in these objects.

## Meteorite and Measurement Techniques

### (A) Analytical Methods

Back-scattered electron images, X-ray elemental abundances and quantitative analyses were obtained using a field-emission electron microprobe analyser JEOL 8530F hyperprobe at NASA Johnson Space Center in Houston. An electron beam accelerated at 20kV, focused to a spot size of 1 micron and carrying a current intensity of 40nA was rastered over the meteoritic sample to obtain characteristic X rays. X-ray intensities and elemental abundances were obtained using a ThermoElectron ultradry SDD (silicon drift detector) energy dispersive spectrometer. Quantitative analyses were performed using a 20nA, focused (1μm) electron beam in wavelength dispersive mode. Five available spectrometers were calibrated using appropriate standards and analogous terrestrial minerals hibonite, plagioclase, sitkhin anorthite were measured following standard procedures prior to the analyses of mineral phases in hibonite-pyroxene spherule. Matrix correction ($\Phi\rho Z$) scheme was applied to quantify the abundances.

### (B) Petrography

Allan Hills (ALHA) 77307 is one of the most pristine samples available in the meteoritic collection and has been classified as a type 3.03 (Bonal et al. 2007; Kimura et al. 2008). About ~181.3 grams of the extraterrestrial rock was collected from the Allan Hills region in Antarctica during the joint US-Japan Antarctic expedition of 1977-78. National Institute of Polar Research, Tokyo kindly provided a thin section of the ALHA77307 meteorite for the study. Previous petrographic, isotopic studies of different components of the meteorite suggest scarce and limited alteration of any of its components (Bonal et al. 2007; Kimura et al. 2008) with a peak metamorphic temperature seen of ~400 °C (Huss et al. 2004). The weathering grade classification of this meteorite to the lowest grade 'Ae' implies it has seen minimal terrestrial alteration (index e referring to the exterior regions only) during its residence on the Antarctic ice fields. Unfortunately, a strong jet of air from the air-dust remover can found way between a small ripped/ detached portion of the meteorite and the glass base plate resulting into scrapping off of the entire meteorite piece off the glass plate. The torn off meteorite section was blown into the ventilation of the laminar flow bench, and most of the thin section was lost prior to any radiogenic isotopic systems ($^{26}$Al-$^{26}$Mg, Be-B), stable isotopes (oxygen, calcium, titanium), and rare earth elements (REE) abundances studies. Thus, only useful petrographic and elemental compositions information of the HIXs are presented and discussed in this paper with the unfortunate limitation of any isotopic data.

The hibonite-pyroxene spherule (RM-1) found in the ALHA77307 is a spherical object with an exposed surface area of about ~120μm in diameter (Fig. 1). Two larger hibonite-pyroxene spherules have been previously reported in Colony (SP1 ~170μm) and ALH 82101 (SP15 ~140μm) (Russell et al. 1998). It has four fragmented hibonite laths that are positioned centrally within the object. Two large hibonite laths of ~

30×15 μm are surrounded by aluminous pyroxene with slivers having sub-micron sized grains. Fe-, Ni metal, and FeS rich grains subsumed in an unidentified Mg-rich phase constitute the slivers towards the two opposite sides of the outer rims. The hibonite show pale blue pleochroism; the pyroxene grain appears colourless with nearly uniform extinction. The two fragmental pieces of hibonites near the two other larger central laths fit the broken edges of the larger hibonite laths (Fig. 1, Fig. S2 see dotted lines around the fragmented portion 'now' and the inferred 'original' position). The edges of the euhedral/subhedral hibonite prismatic crystals along the longer axis (c-axis) are sharp while the two fragmented smaller hibonites seem to be the chipped/broken off part of the larger laths. A reconstruction of hibonite laths prior to break up is shown in Fig. S2 and is facilitated by the prismatic crystal structure and presence of only two hibonites. A similar reconstruction for HIXs in E4631-3 is also shown. Presence of several anhedral hibonites with round edges, do not allow such a reconstruction for any other HIXs. However, along the shorter axis on the lower side of the hibonite laths the sharp cleavage planes are absence and (in Fig. 1) this feature is made ambiguous. Hibonite laths in HIXs within Lancé, E4631-3 show remarkable morphological similarity in size, cleavage plane, occurrence within the object with ALHA77307 compared to others HIXs in Murray and Murchison where there are presence of several fragmented, jagged hibonite scattered randomly within the al-rich pyroxene.

The pyroxene surrounding the hibonites in these, so far found nine objects, span textures ranging from glassy (Lancé 3413-1/31, MUR 7-228) to devitrified glass (Colony SP1, SP15; E4631-3) to crystallized pyroxene grains (MYSM3, Y17-6). The earliest identification of these unique objects in Lancé (Kurat 1975) was serendipitous, piqued by esoteric combination of hibonite laths in glassy pyroxene. It was followed by identification amongst refractory objects obtained by freeze-thaw/ digestion method where in these have been individually separated and collected from Murchison and Murray (Ireland et al. 1991; Simon et al. 1998). Expectedly, HIXs found in the present study has broad morphological similarity with other hibonite-pyroxene spherules, the closest being the HIXs in Lancé (CO3.5) (Fig. S1). However, it is almost 2.5 times bigger in size (diameter ~120μm) than the one in Lancé but the nested hibonite laths in both are of similar size (30×15μm). It also has quite morphological similarity with HIXs in E4631-3 where the pyroxene seem to have been devitrified and/or (re)crystallized to a greater degree.

The outer boundary of the HIXs in ALHA77307 has smooth circular contact (Fig. 1A) quite similar to the Lancé and unlike HIXs from Y 791717 and ALH 82101 (Fig. S1). The HIXs is separated from the matrix by a circular fracture around the object. It also has a few cracks within the pyroxene glass. The circular fracture and cracks are likely to be the result of thin section preparation as the surrounding loosely bound fine- grained matrix material was lost. Fine grained fayalitic matrix material along with fragmented olivine, pyroxene, Fe,-Ni metal, sulphides grains and other CAIs are present in the close proximity to the spherule (Fig. 1B). There is greater abundance of matrix material in an (exposed 2D) area of ~250 μm around HIXs in ALHA 77307 compared to other areas in the thin section. There exist only one additional such 'matrix rich' pocket in the thin section. Most CAIs and chondrules in ALHA 77307 have 'accretionary rims' of sizes 20-150 μm that often also have a continuous outer rim of Fe- rich metal or sulphide grains. Such an 'accretionary' rim around HIXs in ALHA 77307 is conspicuous.

### (C) Mineralogy

**Hibonite: Homogeneous composition and unmelted texture**

The X-ray (Kα) elemental maps of Al, Mg, Ca, Ti, Si are shown in Fig. 1. C-G. The elemental distributions within the object display sharp boundaries. For example, Al, Ca, Si maps show sharp delineation between hibonite laths and surrounding pyroxene. The hibonites have uniform composition with abundance of $Al_2O_3$, MgO, CaO, and $TiO_2$ of 89.5, 0.8 (±0.1), 8.5(±0.1), and 1.8 (±0.1) wt. %, respectively (Fig.1, 2, 3 Table 2). The observed composition of hibonite laths is similar to other hibonite pyroxene spherules, found previously. The hibonite compositions ($Al_2O_3$ 88-90%, CaO ~8.5%, $TiO_2$ 1.5-2.5%) of hibonite-pyroxene spherules are quite similar and seem to be independent of the composition of their surrounding Al-rich pyroxene. The deviation from ideal composition of hibonite of $CaAl_{12}O_{19}$ is commonly observed in meteorites because of coupled substitution of $Mg^{+2}$ and $Ti^{4+}$ for $2Al^{3+}$. Theoretically, titanium concentration ($TiO_2$) up to ~8 wt.% can been incorporated in the crystal structure (Allen et al. 1978) and has been observed in CAIs in various groups of carbonaceous chondrite (MacPherson 2014). However, the titanium ($TiO_2$) abundances within hibonite laths in HIXs show a limited range of 1-2.5 wt.%, in particular in ALHA77307 it is <1.8 wt.%. During the quantification, titanium was assumed to be in the quadrivalent state although some of it may also be present as trivalent charged state (Fig. 4). Presence of about ~12% of total titanium cations as trivalent cation can explain the observed excess in (Si+Ti) cation and the expected 1:1 correspondence with Mg cation. Spinel and perovskite are ubiquitous in CAIs and found abundantly in various petrogenetic settings at times also associated with or within hibonites. However, they have only been found previously in HIXs from Murchison 7-753 E4631-3, and ALH 85085 (Grossman 1988; Ireland et al. 1991; Guan et al. 2000). Two sub-micron perovskite grains are hosted within hibonite in ALHA77307. A representative quantitative profile of elemental abundance (Fig. 1, line a) in hibonite is shown in Fig. 2. A positive correlation between MgO and $TiO_2^{tot}$ can also be seen in hibonite in HIXs. Although, showing a limited range in $TiO_2$ of 1.5 to 1.9, it correlates with abundance of MgO ranging from 0.7 to 1.1. A similar linear correlation over wider range of MgO and $TiO_2$ has been observed for Y17-6 and MYSM3 (Simon et al. 1998).

## Ca, Al- rich pyroxene

Two representative elemental abundance profiles, along two transects in the pyroxene, indicated in Fig.1A are shown in Fig. 2 (lines b and c). Line b profiles a traverse across the entire homogeneous region to the opposite diametrical end while line c follows the Mg enriched region at the boundary of the pyroxene with the sliver zone (Table 3). The line profile b shows general homogeneity (over ~100 μm) along the entire stretch. At the outer boundary regions, however a slight enrichment in $Al_2O_3$ (~2.0 wt. %), which is compensated by combined lower abundance of $SiO_2$ and MgO, is observed (Fig. 1, Fig. 2). The core region has CaO and $Al_2O_3$ abundances of ~ 25 and ~30 wt. %, respectively. The inverse correlation of $Al_2O_3$ with MgO and $SiO_2$ can be seen. The CaO abundance however is markedly uniform. The line profile c traverses the region around the sliver zone. A significant variation in abundance of $Al_2O_3$, MgO and $SiO_2$ over a short distance (5-10μm; entire length of profile is ~20μm) of can be noted. The expected inverse correlation between the abundances $Al_2O_3$ with MgO and $SiO_2$ of these resulting from crystallization of pyroxene is inferred form the Fig. 2. The Al-rich pyroxene surrounding the hibonite shows a mild zonation in Al, Ti with marginally Al-rich composition towards the core and enriched Ti concentration in the outer regions. The correlations between these elemental abundances within pyroxene in HIXs have been previously noted (See, Fig. 3, 4 of Simon et al., 1998).

**Other Accessory phases in the wedges**

Several phases are present at the outer margins of the inclusion and consist of a mélange of poorly/rapidly crystallized grains. These phases in the sliver zones represent a minor component of the HIXs. However, their textures, location, and mineral compositions provide significant important information about the formation and alteration of HIXs. The mushy sliver zones host submicron Fe, Ni metal grain, FeS, and an unidentified Mg-, Ca-, Si bearing phase. Per se, as noted previously, an important accessory phase found hosted within the hibonite is pervoskite, which is present as two sub-micron grains. These sub-micron phases were identified from the EDS spectra.

**Discussion**

**Morphological Similarity and Origin of Diversity Amongst Hibonite-Pyroxene Spherules**

Of the ten hibonite-pyroxene spherules reported so far, nine have been reported in carbonaceous chondrites with five of those found in CO chondrites of varying petrologic types (3.0-3.5). These objects are characterized by their defining unique mineralogy and shape. Notwithstanding this morphological and mineralogical commonality, there exist important differences. Some of the important observed differences could emanate from differences in their initial bulk elemental and isotopic compositions (Ca, Ti isotopes, $^{26}$Al abundance) and conditions of formation, while others could be superimposed effects of nebular and parent body metamorphism. It is rather difficult to decouple the effects of nebular metamorphism from parent body metamorphism. The differences are difficult to deconvolve owing to significant variations amongst meteorites in the amount of bulk $H_2O$ (water ice) incorporated, peak temperatures, porosity, size of parent body, cooling rates etc.

Classification of unequilibrated chondrites into petrologic type 3.0-3.9 is based on increasing level of thermal metamorphism experienced by them which results in characteristics shifts in several properties (e.g., texture, cathodoluminiscence of mesostases glass; abundance of presolar grains in matrix, deviation in $Cr_2O_3$ abundances in olivine; maturity of insoluble organic matter etc.) of various components of the meteorites (Huss et al. 2004). Some of the observed differences in five HIXs in CO chondrites of different petrologic type could be attributed to the varied level of thermal metamorphism. The HIXs in the least altered meteorite ALHA77307 is therefore important because it provides a case wherein only limited parent body metamorphism is expected and is observed. Its morphology and mineralogy hence provides the end member. With the above mentioned caveats and the absence of anorthite, melilite in hibonite-pyroxene spherules, that are more amenable to alteration, the range of morphological textures seen in the hibonite-pyroxene spherules can be set according to their metamorphic grade, with ALHA77307 defining the pristine texture. The original typical morphology was overprinted/ altered to variable degree corresponding to their metamorphic grade resulting in glassy to crystalline textures (Fig. S1). While such a characteristic trend is expected amongst HIXs, it was obfuscated by significant, most probably, parent body metamorphism of the Colony and ALHA 82101. Until the present study Colony, with petrographic grade of 3.1, was the lowest grade meteorite where hibonite-pyroxene spherule has been reported. The importance of the HIXs in ALHA77307 lies in

observation/ finding the pristine character that allows to discern a general trend of progressive overprinting/ alteration of the morphological character and also possibly isotopic records.

## Formation Scenario of Hibonite-Pyroxene Spherules

Inferences from the Morphology and Mineralogy

Common Genesis scenario: Simon et al. (1998) inferred that hibonite-pyroxene spherule form by aggregation/entrapment of hibonites within a rapidly cooling Al-, Ca-rich pyroxene melt. The maximum temperature experienced and cooling rate of the melt varies between HIXs. However, the maximum temperature could be high enough for sufficiently long time duration to melt hibonite and equilibrate partially, or completely with the melt. The melt is cooled sufficiently fast such that anorthite crystallization is suppressed and varied textures (glassy, crystalline) with different degree of zoning pattern is seen in these objects. The texture and elemental abundances in ALHA77307 spherule (RM-1) is consistent with the posited rapid cooling scenario (Simon et al. 1998). The hibonites laths in HIXs in ALHA 77307 have mostly sharp edges and are devoid of rounded boundaries or embayments except for along one end of the shorter axis of the exposed surface. The textural features along with the elemental concentration profiles suggest that there was no, or at worst only minimal, interaction between the hibonite and pyroxene. The two fragmental smaller pieces of hibonites fit rather well the two larger hibonite laths. It allows to infer that (i) some shock event broke the pre existing hibonite laths just before aggregation into the pyroxene melt (ii) the melting temperature were either not high enough to melt hibonites and/or of short duration to leave behind any evidence of equilibration between hibonite and pyroxene (iii) hibonite and pyroxene did not crystallize from the same melt. Hence, hibonite laths in ALHA 77307 HIXs are most likely relict. The range of minimum and maximum temperature can be inferred from the melting point of Ca-pyroxene and hibonite. But the melting point of hibonite depends on the total pressure and hence an exact determination is not possible in the present study. Isotopic studies (oxygen, $^{26}$Al-$^{26}$Mg, REE) could have firmly established the assertion. The lack of isotopic studies prohibits stricter constraint. The higher abundances of Al, and Ti at the outer boundaries of pyroxene imply that pyroxene crystallized from outwards. Presence of sliver zones at the outer edges of the pyroxene and absence of anorthite are inferred to mean that cooling rate was high and consumed Mg to leave behind a slightly Al-enriched, Ca-pyroxene to crystallize in the central region of the HIXs (Fig. 1 A-G, Fig. 2 Line b).

One of the most salient features of the HIXs in ALHA77307 is the presence of unaltered phases in the slivers at the outer regions. The region identified, as sliver zone in ALHA 77307 constitute a significantly greater part in ALHA 82101 and Y 791717. They are less prominent in the smaller Lancé inclusion (See Fig. 1 of Ireland et al. 1991). The forsterite matrix like material adhering to the HIXs, obtained by freeze thaw/ disaggregation method, could be part of such zone (Ireland et al. 1991; Russell et al. 1998; Simon et al. 1998). The minerals consisting of Fe, -Ni, metal, FeS in sliver zones of HIXs in ALHA 77307 are irregularly shaped, poorly crystallized and randomly oriented. Such a texture could form in melt inclusion kind of texture by assimilation of most of the elements incompatible to Al-rich pyroxene melt. The incompatible elements making the initial bulk composition of the Al-rich, Ca-pyroxene melt were sequestered into wedged slivers/ melt inclusion kind of texture at the outer regions of this HIXs. The matrix surrounding the HIXs in ALHA77307 contains fine grained fayalitic olivine, Fe,-Ni metal grains, FeS etc.. The mushy, irregular textured grains in the slivers are devoid of any connecting melt/vein and lack any mineralogical correlation with the surrounding matrix. These textural features and the very low petrologic type (3.03) of the meteorite suggest that these phases in the sliver zones are the representative last dregs of the melt and not alteration products. Alternatively, the mineral phases in the sliver zone could be interpreted as the trapped matrix material. This interpretation is not favored because (1) the melting of Ca-pyroxene should have also melted the matrix material and got assimilated into the bulk composition of the melt (2) similar regions of larger dimensions are observed in other HIXs. Crystallization of pyroxene leads to zoning in elemental abundance distributions and expected relationship between $Al_2O_3$, $MgO$, $TiO_2$, and $SiO_2$ (Simon et al. 1998). The zoning of pyroxene is a differentiating characteristic between the glassy and crystallized pyroxene. Slower cooling or later stage heating should typically result in formation of several micro-crystallites and/or bladed acicular Mg rich grains in the sliver zones (See, for e.g., chondrule#1, 3 of Semarkona and chondrule#1 of Vigarano (Mishra and Goswami 2014; Mishra and Chaussidon 2014b) chondrule #5 in QUE 97008 (Mishra et al. 2016)). Such a texture resulting from slightly slower cooling and/or later stage reheating is seen in HIXs in ALH 82101 and Y17-6. The pyroxene in HIXs (RM-1) in ALHA 77307 shows zoned character. The central (80-100 μm) region around the hibonite laths is quite homogeneous and the zoning pattern in Al, Mg, Si, Ti becomes apparent towards the peripheral 20-30 μm. The expected relationship between elements resulting from crystallization of Al-rich pyroxene can be seen in Fig. 1-3. Simon et al. (1998) have previously studied two hibonite-pyroxene spherules from Murray (CM2, MYSM3) and Yamato 791717 (CO3.3) and extensively discussed petrogenesis of those objects. There are several similarities in petrography and mineralogy between HIXs studied by Simon et al. (1998) and HIXs in ALHA 77307. However, both the HIXs studied by Simon et al.

(1998) are more aluminous and less siliceous (~ 2 wt. %) in the calculated bulk (Fig. 5, Table 4). While Y17-6 is of similar size as that of HIXs in ALHA 77307, the one is Murray (MYS3) is only half as large. The correlation trends between elemental abundances resulting from crystallization of pyroxene are seen over a wider range of elemental redistribution in pyroxenes in Murray (MYSM3) and Y 791717 compared to that in HIXs from ALHA 77307 (Fig. 3). The difference can be understood in terms of faster cooling for HIXs in ALHA 77307 compared to the other two HIXs. A smoother texture of pyroxene also seems to be consistent with the observation of faster cooling. Alternatively, it could be explained by difference in the bulk composition or the difference in the dust enrichment the local environment (density of particles/gas, maximum heating temperature, cooling rate etc.) in which these objects formed.

**Evidence of Alteration amongst HIXs**

The hibonite-pyroxene spherules in different meteorites belong to different petrographic grade and hence have seen varied level of aqueous and thermal metamorphism. The varied level of metamorphism is also reflected in the mineral compositions of the accessory, alteration phases in these spherules. HIXs reported in the present study shows a marked pristine character quite consistent with its petrologic type. Most of the spherules studied previously were found to be associated with some phases that were interpreted to have terrestrial or metamorphic related origins. Veins of Fe-rich material from terrestrial weathering were found cross-cutting and surrounding the hibonite-spherule in Colony (CO3.1) (Russell et al. 1998). In ALH 82101 (CO3.3) spherule the 'sliver last dreg zone' at the edge of the inclusion has been altered to fine grained feldspathoid and pyroxene (Russell et al. 1998). An observation of presence of nepheline in a morphological similar zone was noted within the Y17-6 spherule (Simon et al. 1998). Forsteritic material with several cracks was also noted in MYSM3. The analogous region in Lancé also has most probably been significantly altered to rather bear resemblance with the surrounding matrix and lose the perspective associative character with the HIXs (Fig. S1 Note the lower edge of HIXs in Lancé; Ireland et al. 1991). Additionally, the outer edges of 'HIXs'es in Colony, ALHA 82101, Y17-6, EET 87746, ALH 85085, and Lancé have been corroded to have uneven, jagged outer most boundary (Fig. S1). Such clearly visible distinctive marks of post formation interactions and parent body thermal metamorphism are absent in the HIXs from ALHA77307 and cannot be inferred for other HIXs which were obtained following freeze-thaw procedure. Elemental abundance maps of Na, S, Fe, do not show any obvious relationship between the HIXs (RM-1) and surrounding matrix. Thus, to first order it can be inferred that 'last dreg zone' containing fine, small-grained minerals was more easily altered to varying degrees, consistent with their petrologic type of the meteorites in which they were found.

The average composition of hibonite, pyroxene, and the calculated bulk of the HIXs in ALHA77307 are tabulated in Table 4. For calculating the bulk composition, hibonite and sliver zone were considered to make up about 10% and 1%, respectively while the rest of the inclusion being pyroxene. The exposed 2D surface area of hibonite to area of the HIXs is ~10% and was taken as the representative for the bulk. For comparison bulk compositions of some of the other HIXs are also shown. The calculated bulk composition is expectedly similar to other HIXs. The average pyroxene and calculated bulk is plotted on a spinel projection CMAS diagram (Fig. 5) (Huss et al. 2001; MacPherson 2014). Under equilibrium conditions, the bulk composition therefore suggests formation of anorthite prior to formation of pyroxene. This observed inconsistency with the mineralogy of HIXs has been explained previously by invoking rapid undercooling, the absence of nucleating sites, and various rates of fast and slow cooling (0.5 °C/h- 5°C/h). The scenario can also explain the formation of spherule in ALHA77307.

## Summary and Conclusions

A rare kind of refractory object hallmarked by hibonite lath surrounded by Al-rich pyroxene, defined previously as Hibonite-Pyroxene Spherule, has been found in one of the most pristine meteorite Allan Hills 77307 (CO3.03). It provides the end member characteristics in term of mineralogy and petrography and thereby allows to the discernment of a plausible general trend of formation and subsequent partial recrystallization and progressive alteration with increasing petrologic type. The petrography and mineralogy supports its formation by rapid cooling of the pyroxene without significant melting/ interaction of the internal hibonite laths. Previous proposed scenario of formation of these objects by entrapping of a relict hibonite laths within a fast cooled Al-rich pyroxene melt can explain formation of the hibonite-pyroxene spherule (RM-1) in ALHA77307.

## Acknowledgements

I'm grateful to National Institute of Polar Research, Tokyo, for lending a thin section of ALHA77307. My sincerely thank Dr. D.K. Ross (Johnson Space Center, Houston) for providing generous assistance during mapping and quantitative analyses. I also acknowledge several fruitful discussions with Marc Chaussidon, Justin Simon, Kuljeet K Marhas, Daniel K Ross, Jangmi Han, and Alejandro Cisneros. The financial support during the work by NASA post-doctoral program (NPP) fellowship at NASA Johnson Space Center, Houston is gratefully acknowledged. The manuscript was prepared during the Humboldt fellowship at Heidelberg University. Financial support from the Alexander Von Humboldt foundation during the term is also sincerely acknowledged.

Table captions:

Tables:

Table 1.

General features of the hibonite-pyroxene spherules

Table 2.

Electron microprobe analyses of hibonite

Table 3.

Electron microprobe analyses of pyroxene

Table 4.

Chemical compositions of hibonite, pyroxene, and calculated bulk of hibonite-pyroxene spherules

Figure captions:

Figures:

Fig. 1

Back-scattered electron image (A), X-ray elemental abundance maps of Al, Mg, Ca, Ti (C-F) and (B) a composite map (Mg-Ca-Al as RGB) of the hibonite-pyroxene spherule in Allan Hills 77307 (CO3.03). The scale bar is shown in BSE image.

Fig. 2

Representative EPMA quantitative linear profiles, indicated in Fig. 1A, across a hibonite and two transects along the surrounding pyroxene is shown. Note the analyses points along the three linear profiles (a-c) are spaced unequally on the HIXs. The linear profile 'a' ends closer to the edge of hibonite crystals and excitation volume of surrounding pyroxene is attributed for the marginal lower abundance of $Al_2O_3$ in the last 2 data points of the profile 'a'.

Fig. 3

$SiO_2$ correlation diagram in pyroxene of hibonite-pyroxene spherules

The plot shows variation of MgO, $Al_2O_3$, $TiO_2$ (Y axis) with respect to $SiO_2$ (X axis) for analyses within pyroxene. Corresponding ranges observed in pyroxene in Yamato 791717 and Murray MYSM3 (Simon et al. 1998) are shown by light and dark gray filled rectangular boxes, respectively. Open rhombuses show $Al_2O_3$ (upper) and MgO composition of pyroxene glass in Lancé, with respect to $SiO_2$ (Ireland et al. 1991).

Fig. 4

Quantitative analyses of hibonite laths from the ALHA 77307 HIXs and the corresponding relationship between abundance of Mg and (Si+ Ti) cations in hibonites.

Fig. 5

Spinel projection CMAS diagram. The bulk composition of the analysed hibonite-pyroxene spherule from ALHA77307 along with previously studied HIXs, is projected from spinel ($MgAl_2O_4$) onto the plane $Al_2O_3$-$Mg_2SiO_4$-$Ca_2SiO_4$ following Huss et al. (2001) and MacPherson (2014). Black lines show experimental phase boundaries. Condensation trends of solid phases from a gas of Solar composition are shown by the red lines. Abbreviations are Cor for corundum, Hib for hibonite, An for anorthite, Ak for akermanite, Gro for grossite, Geh for gehlenite, Di for diopside, En for enstatite, La for larnite, Mw for merwinite, Mo for monticellite, Fo for forsterite.

Fig. S1 Back scattered electron images of Hibonite-Pyroxene Spherules.

HIXs found in (A) ALHA 77307 in the present study (B, C) in Colony and ALH 82101 from Russell et al. 1998 (D-F) in Lancé and Murchison from Ireland et al. 1991 (G) in EET 87746 from Guan et al. 2000 (H, I) in Murray and Yamato 791717 from Simon et al. 1998. Scale bars are shown in each panel. Except for hibonite-pyroxene spherule from Yamato 791717 others have been arranged in the order of increasing petrologic type. Images have been taken from published literature. Note the trend of increasing crystallinity of pyroxene, corrugation of the outermost boundaries and alteration features in the hibonite-pyroxene spherules.

Fig. S2

Inferred original structure of hibonites in Hibonite-Pyroxene Spherules

in (A) ALHA 77307 (this study) (B) EET 87746 (Guan et al. 2000).

Thicker dotted lines show the positioning of the fragmental pieces (A, B, C) of hibonites prior to breaking up. Thinner lines are drawn around the fragmented pieces (A`, B`, C`) of hibonites. Some space has been deliberately left between the fragmented pieces to clearly show the edges of both the pieces. Arrows indicate the direction of motion. A dust particle was sitting at the edge of the hibonite grain in the X-ray elemental abundances maps shown in Fig. 2. It is absent in the image here. The smooth outer margin of HIXs and absence of petrographic correlations between the grains in the sliver zones and the grains in the outer matrix can be seen clearly.

Fig. 1

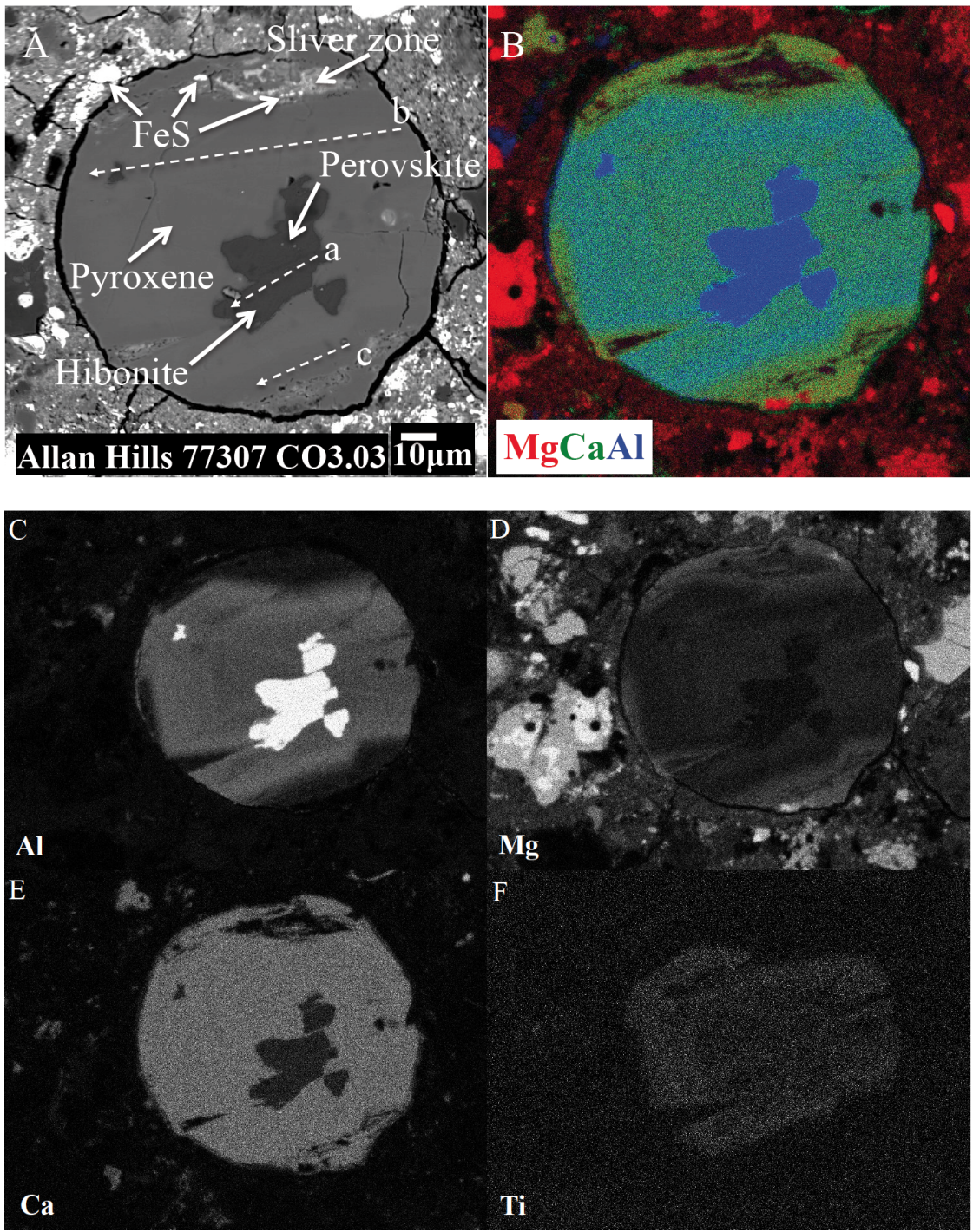

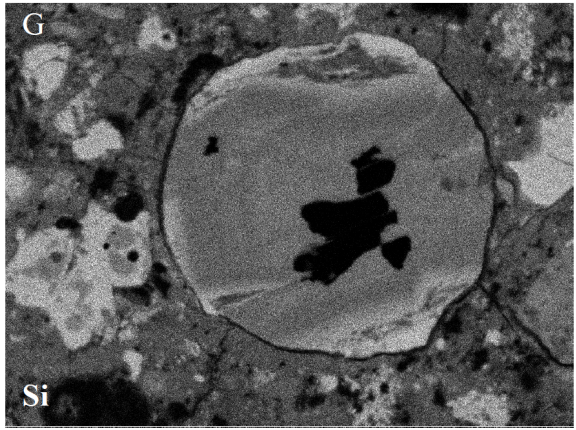

Fig. 2

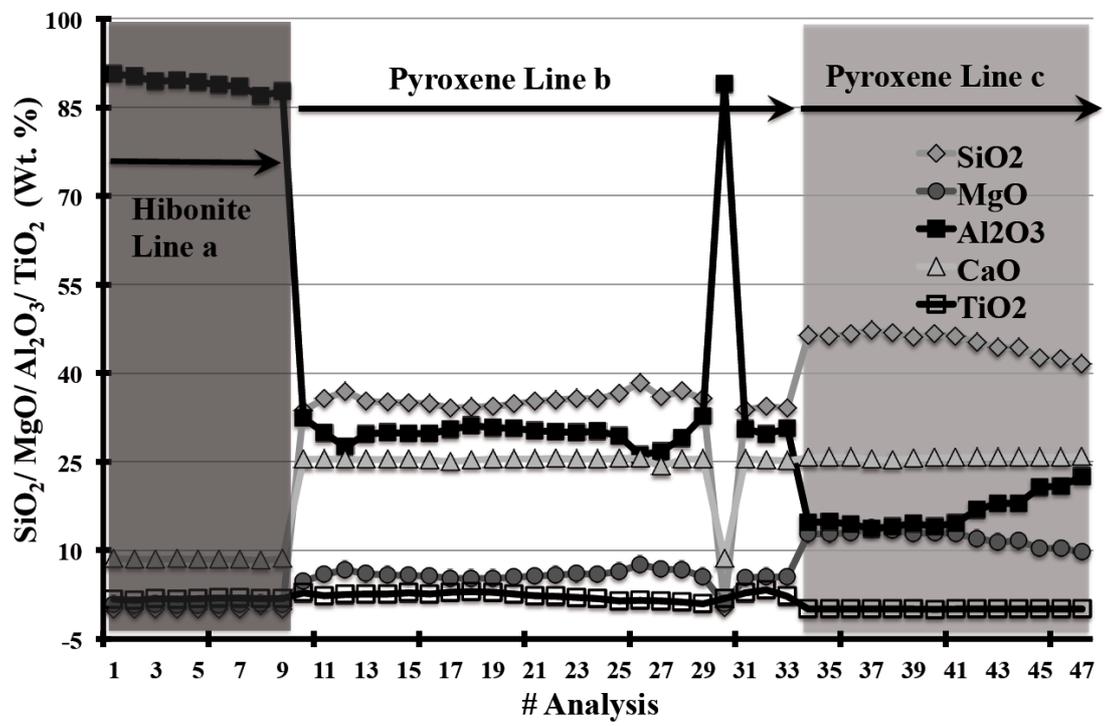

Fig. 3

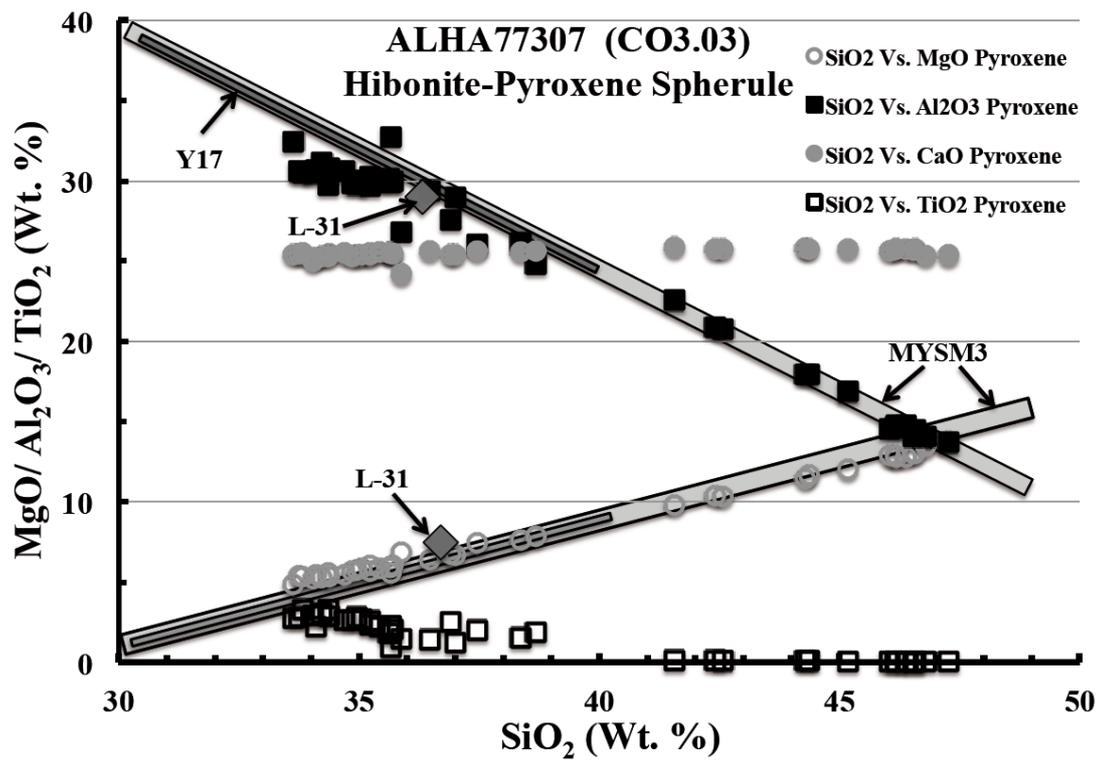

Fig. 4

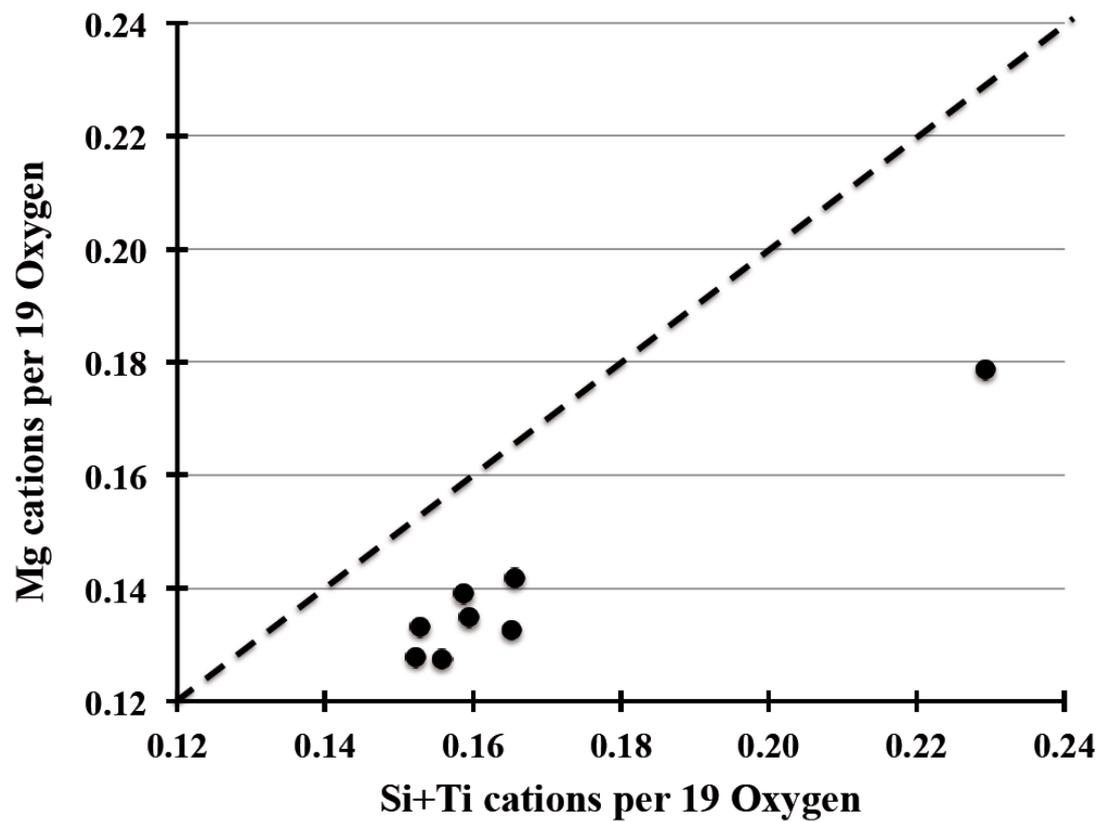

Fig. 5

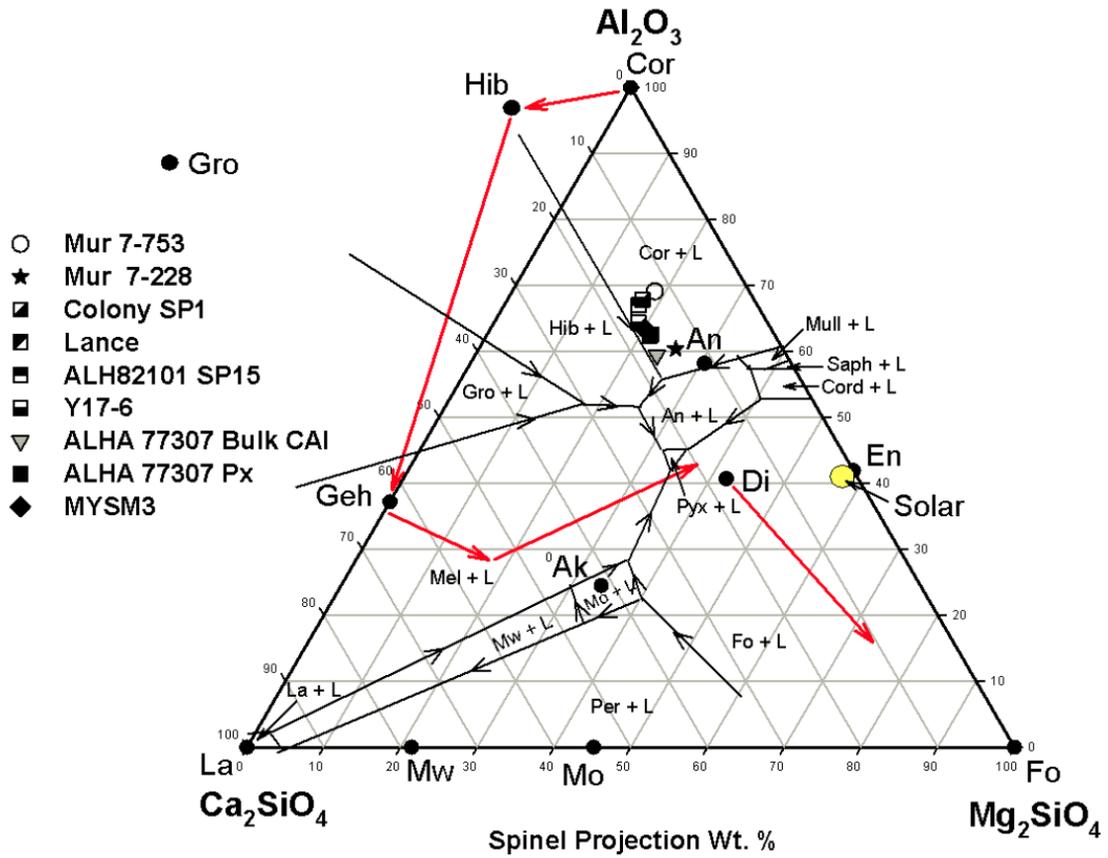

Fig. S1

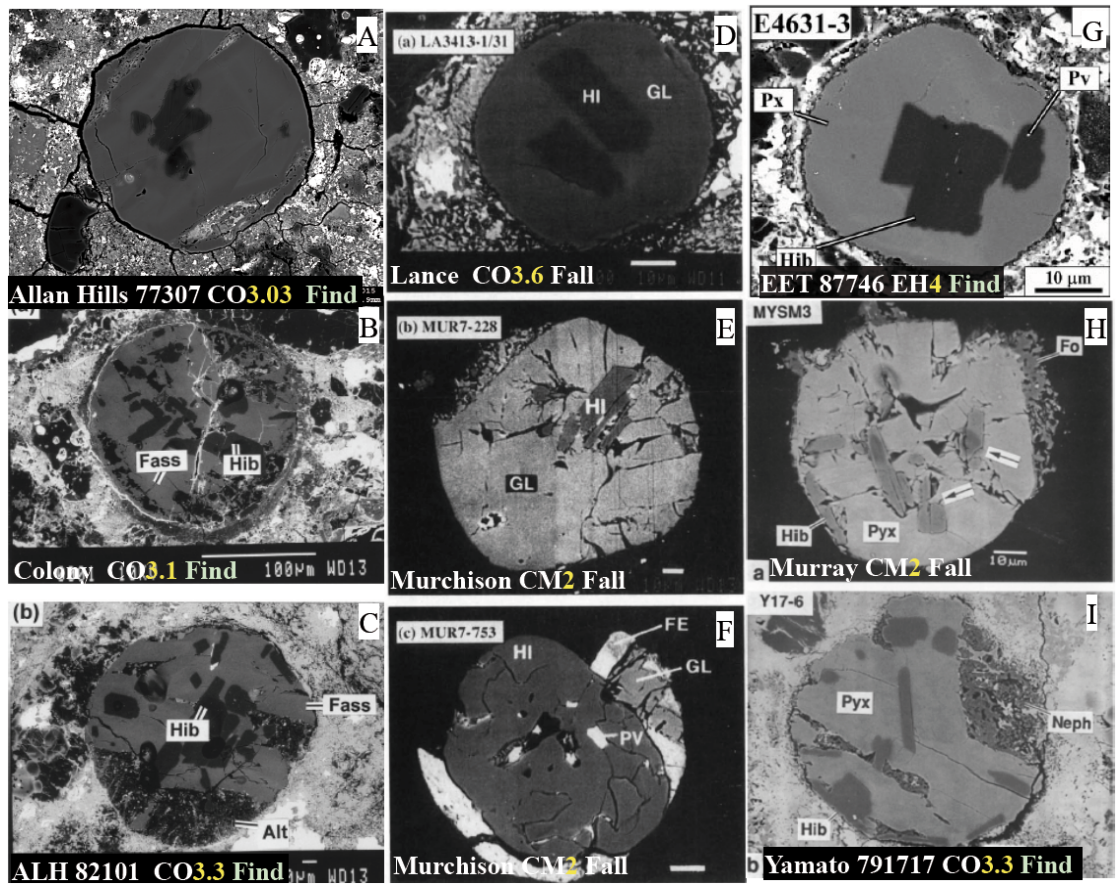

Fig. S2

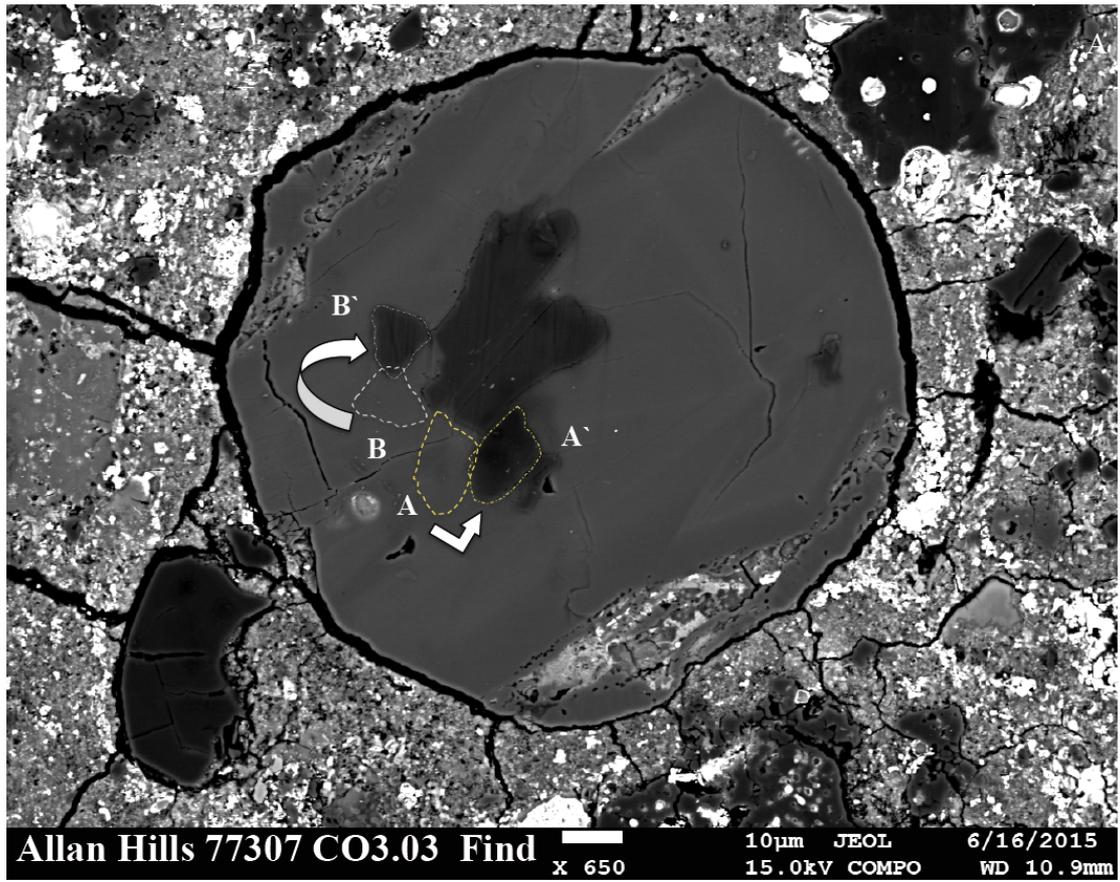

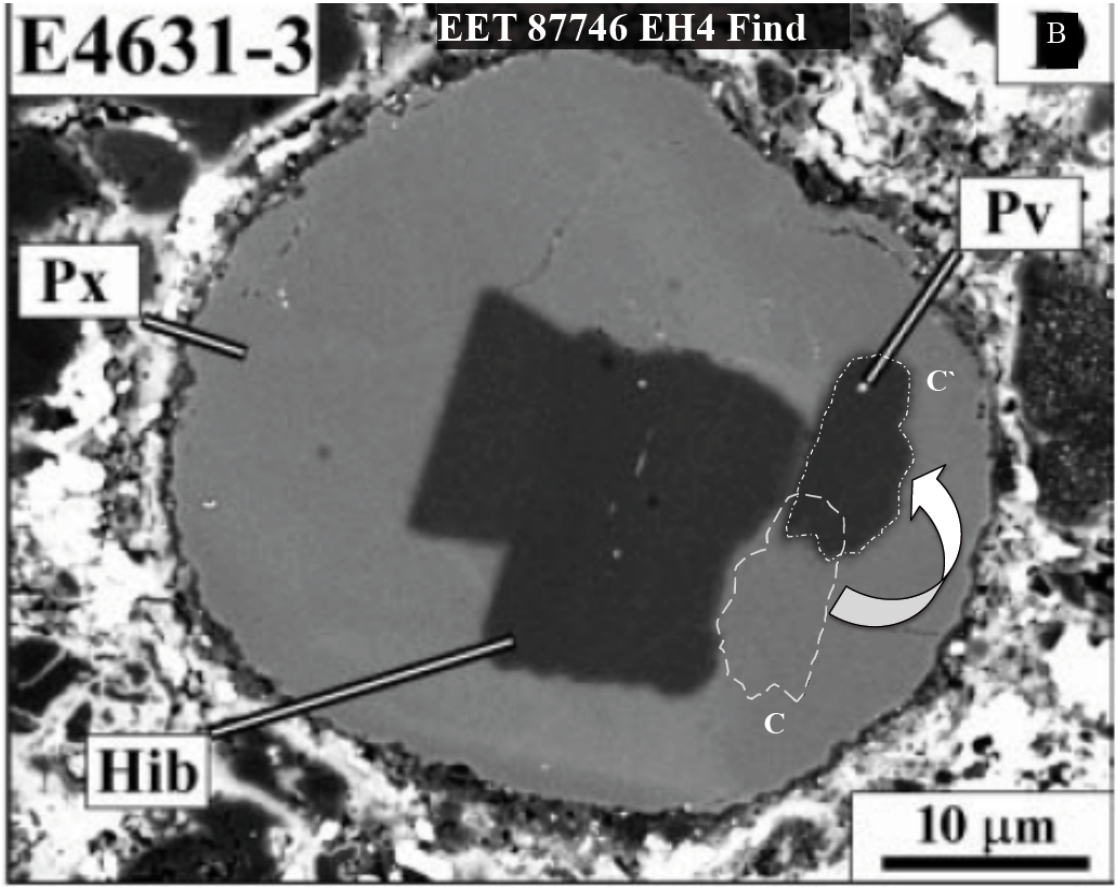

Table 1. Properties of Meteorites and features of hibonites-pyroxene spherules

| Meteorite | Sample Identifier | Petrologic Type | Fall/ Find | Year found | Weathering grade | Total Mass of Meteorite | Dimensions HIXs (μm) | Hibonite lath (μm) | Texture of Pyroxene | Zoning in Pyroxene | $^{26}Al/^{27}Al \pm 2\sigma$ $(10^{-6})$ | F ($\delta^{25}Mg$ ‰) in Pyroxene | REE pattern hibonite | Reference |
|---|---|---|---|---|---|---|---|---|---|---|---|---|---|---|
| ALHA 77307 | RM-1 | CO3.03 | Find | 1977 | A | 181.3 g | 120×120 | 30×15 | Crystalline | yes | nd | nd | nd | 8 |
| Colony | SP1 | CO3.1 | Find | 1975 | - | 3.91 kg | 170×170 | 70×25 | Devitrified | no | UR | - | II | 5 |
| Y 791717 | 17-6 | CO3.3 | Find | 1979 | - | 25.32 kg | 120×120 | 40×5 | Crystalline | yes | 200-300* | -6.2 | II | 3,6 |
| ALH 82101 | SP15 | CO3.4 | Find | 1982 | - | 29.1 g | 140×140 | 60×50 | Devitrified | no | 4.4±3.4 | - | II | 5 |
| Lance | 3413-1/31 | CO3.5 | Fall | 1872 | n/a | 51.7 kg | 50×50 | 20×10 | Glassy | no | UR | 8.2 | II | 1,4 |
| Murchison | 7-228 | CM2 | Fall | 1969 | n/a | 100 kg | 120×120 | 40×15 | Glassy | no | 17±7 | 9.2 | Frac, Enrich LREE | 4 |
| Murchison | 7-753 | CM2 | Fall | 1969 | n/a | 100 kg | 70×68 | 50×68 | Glassy | no | UR | -3.2 | Unusual | 4 |
| Murray | MY92S3 | CM2 | Fall | 1950 | n/a | 12.6 kg | 70×60 | 25×10 | Crystalline | yes | UR | -3.4 | II? | 6 |
| ALH 85085 | F-23 | CH3 | Find | 1985 | A/B | 11.9 g | 11×11 | 5×1.5 | Glassy | no | nd | nd | nd | 2 |
| EET 87746 | 4631-3 | EH4 | Find | 1987 | C | 142.3 g | 40×40 | 15×10 | Crystalline | ? | 2±5 | -5.3 | nd | 7 |

*In glass and no excess in hibonite [1] Kurat, 1975; [2] Grossman et al. 1988; [3] Tomoeka et al. 1992; [4] Ireland et al. 1991; [5] Russell et al. 1998; [6] Simon et al. 1998; [7] Guan et al. 2000; [8] present study. UR : unresolved excesses; nd: not determined

Table 2. Electron microprobe analyses of hibonite

|  | 1 | 2 | 3 | 4 | 5 | 6 |
|---|---|---|---|---|---|---|
| $SiO_2$ | 0.04 | 0.04 | 0.04 | 0.04 | 0.04 | 0.06 |
| $MgO$ | 0.82 | 0.77 | 0.77 | 0.80 | 0.85 | 0.83 |
| $Al_2O_3$ | 89.34 | 89.59 | 89.27 | 88.78 | 88.48 | 87.68 |
| $CaO$ | 8.49 | 8.55 | 8.49 | 8.52 | 8.50 | 8.62 |
| $TiO_2^{tot}$ | 1.86 | 1.78 | 1.82 | 1.92 | 1.92 | 1.79 |
| $FeO$ | 0.14 | 0.12 | 0.16 | 0.14 | 0.14 | 0.16 |
| Total | 100.69 | 100.84 | 100.55 | 100.20 | 99.93 | 99.14 |
| Cations per 19 oxygen anions | | | | | | |
| Si | 0.005 | 0.004 | 0.004 | 0.005 | 0.005 | 0.007 |
| Mg | 0.135 | 0.128 | 0.127 | 0.133 | 0.142 | 0.139 |
| Al | 11.680 | 11.693 | 11.686 | 11.669 | 11.661 | 11.653 |
| Ca | 1.009 | 1.015 | 1.010 | 1.018 | 1.019 | 1.041 |
| Ti | 0.155 | 0.148 | 0.152 | 0.161 | 0.161 | 0.152 |
| Fe | 0.013 | 0.011 | 0.015 | 0.013 | 0.013 | 0.015 |
| Total | 12.996 | 12.998 | 12.994 | 12.997 | 13.000 | 13.007 |

Table 3. Electron microprobe analyses of pyroxene

|  | 1 | 2 | 3 | 4 | 5 | 6 |
|---|---|---|---|---|---|---|
| $SiO_2$ | 35.62 | 36.47 | 38.37 | 46.57 | 47.26 | 46.79 |
| MgO | 5.93 | 6.38 | 7.61 | 13.01 | 13.77 | 13.45 |
| $Al_2O_3$ | 30.12 | 29.41 | 26.15 | 14.45 | 13.67 | 14.05 |
| CaO | 25.51 | 25.58 | 25.54 | 25.77 | 25.35 | 25.31 |
| $TiO_2^{tot}$ | 1.80 | 1.41 | 1.53 | 0.03 | 0.03 | 0.03 |
| $Cr_2O_3$ | 0.06 | 0.04 | 0.05 | 0.07 | 0.09 | 0.07 |
| FeO | 0.18 | 0.20 | 0.23 | 0.32 | 0.35 | 0.35 |
| Total | 99.21 | 99.49 | 99.48 | 100.22 | 100.52 | 100.05 |
| Cations per 6 oxygen anions | | | | | | |
| Si | 1.306 | 1.332 | 1.401 | 1.681 | 1.699 | 1.690 |
| Mg | 0.324 | 0.347 | 0.414 | 0.700 | 0.738 | 0.724 |
| Al | 1.302 | 1.266 | 1.126 | 0.615 | 0.579 | 0.598 |
| Ca | 1.003 | 1.001 | 0.999 | 0.997 | 0.976 | 0.980 |
| Ti | 0.050 | 0.039 | 0.042 | 0.001 | 0.001 | 0.001 |
| Cr | 0.002 | 0.001 | 0.001 | 0.002 | 0.002 | 0.002 |
| Fe | 0.006 | 0.006 | 0.007 | 0.010 | 0.010 | 0.011 |
| Total | 3.991 | 3.992 | 3.991 | 4.006 | 4.006 | 4.005 |

Table 4. Bulk compositions of Hibonite-Pyroxene Spherules

| Element Oxide | Allan Hills 77307 RM-1 | | | | Lance 3413-1/31 | Colony SP1 | ALH 82101 SP15 | Yamato 79171717-6 | Murray MYSM3 | Murchison Mur 7-228 | Murchison Mur 7-753 |
|---|---|---|---|---|---|---|---|---|---|---|---|
| | Hibonite | Pyroxene | Pyroxene* | Bulk Calc. | Bulk | Bulk | Bulk | Bulk | Bulk | Bulk | Bulk |
| $SiO_2$ | 0.0 | 37.0 | 46.2 | 32.9 | 33.1 | 26.7 | 27.2 | 29.3 | 31.3 | 39.4 | 27.3 |
| MgO | 0.9 | 6.7 | 12.8 | 6.4 | 6.3 | 5.1 | 5.1 | 5.1 | 6.1 | 7.2 | 5.3 |
| $Al_2O_3$ | 88.5 | 29.0 | 14.6 | 35.4 | 35.1 | 44.9 | 44.0 | 39.3 | 37.4 | 28.9 | 47.0 |
| CaO | 8.5 | 25.4 | 25.7 | 23.3 | 23.8 | 20.1 | 21.3 | 23.1 | 23.5 | 22.6 | 17.9 |
| $TiO_2$ | 1.9 | 1.2 | 0.0 | 1.3 | 1.7 | 2.2 | 3.3 | 3.2 | 1.7 | 1.8 | 2.6 |
| Sum | 99.8 | 99.3 | 99.4 | 99.3 | 100.0 | 98.9 | 100.9 | 100.0 | 100.0 | 99.9 | 100.1 |

Data from relevant literature referred previously in table 1. [4] Ireland et al. 1991; [5] Russell et al. 1998; [6] Simon et al. 1998; * Indicates average composition of (n = 10) pyroxene near the sliver zone.